\documentclass[10pt,paper]{JHEP3}
\usepackage{amsmath}

\title{Higher N-point Amplitudes in Open Superstring Theory\footnote{To be published in
the proceedings of the 5th International Conference on
Mathematical methods in Physics (IC2006), Rio de Janeiro, Brazil,
24-28 April 2006.}}

\author{Ricardo Medina\\
        Univ. Federal de Itajub\'a (UNIFEI), Brazil\\
        E-mail: \email{rmedina@unifei.edu.br}}

\author{Luiz Antonio Barreiro\\
        Funda\c{c}\~ao de Ensino e Pesquisa de Itajub\'a (FEPI), Brazil \\
        E-mail: \email{barreiro@unifei.edu.br}}

\abstract{In this work we report on recent progress in the
calculation of open superstring scattering amplitudes, at tree
level, with more than four external massless states. We also
report on the corresponding terms in the low energy effective
lagrangian.}

\preprint{}


\begin{document}

\section{Introduction and review of known results}

\subsection{The Born-Infeld lagrangian in the context of String Theory}

\noindent As a starting point we consider the low energy interaction of {\it abelian}
open (bosonic) strings in Minkowski spacetime, which is given by the Born-Infeld
lagrangian \cite{Fradkin1}:
\begin{eqnarray}
{\cal L}_{\rm BI} = -\frac{1}{(2 \pi \alpha')^{D/2}}
\sqrt{\mbox{det}(\eta_{\mu \nu} + 2 \pi \alpha' F_{\mu \nu})} \ ,
\label{LBI}
\end{eqnarray}
as long as $F_{\mu \nu}$ is kept constant. $D$ is the spacetime dimension.\\
The $\alpha'$ expansion of ${\cal L}_{\rm BI}$ has the following form:
\begin{eqnarray}
{\cal L}_{\rm BI} = (\mbox{constant}) - \frac{1}{4} F_{\mu \nu} F^{\mu \nu} +
\frac{\pi^2}{2} {\alpha'}^2
\biggl(F_{\mu \nu} F^{\nu \rho} F_{\rho \sigma} F^{\sigma \nu} -
\frac{1}{4} F_{\mu \nu} F^{\mu \nu} F_{\rho \sigma} F^{\rho \sigma} \biggr)
+ {\cal O}({\alpha'}^4) \ .
\label{LBI1}
\end{eqnarray}
The first non constant term clearly is identified as the Maxwell lagrangian and the
${\cal O}({\alpha'}^2)$ term is the first low energy correction coming from String Theory.
Only $even$ powers of $\alpha'$ show up in this expansion.\\
The general situation for the low energy effective
lagrangian includes as well derivatives of the $F$'s:
\begin{eqnarray}
{\cal L}_{\rm eff} = {\cal L}_{\rm BI} + (\mbox{derivative terms}) \ .
\label{Leff}
\end{eqnarray}
The situation for the low energy effective lagrangian in Open Superstring Theory is similar
to the one in (\ref{Leff}), the only difference being the fact that fermionic degrees
of freedom are also present in the lagrangian \cite{Tseytlin1}. Anyway, the bosonic part of this lagrangian has a similar structure as the one in (\ref{Leff}).\\
A $nonabelian$ generalization of (\ref{Leff}) is of interest for Type I theory. A first guess would be to consider the trace of the lagrangian in (\ref{LBI}) and (\ref{LBI1}), leading to
\begin{eqnarray}
{\cal L}^{non-ab}_{\rm BI} & = & (\mbox{constant}) - \frac{1}{4} \mbox{tr} \biggl(F_{\mu \nu} F^{\mu \nu} \biggr) + \frac{\pi^2}{2} {\alpha'}^2 \mbox{tr} \biggl (F_{\mu \nu} F^{\nu \rho} F_{\rho \sigma} F^{\sigma \nu} - \frac{1}{4} F_{\mu \nu} F^{\mu \nu} F_{\rho \sigma} F^{\rho \sigma} \biggr) + \nonumber \\
&& + {\cal O}({\alpha'}^4) \ ,
\label{LnaBI}
\end{eqnarray}
where, clearly, the first non constant term is the Yang-Mills lagrangian. From now on we will
des-\
consider the constant term in (\ref{LnaBI}).\\
The problem arises as soon as we consider the ${\cal
O}({\alpha'}^2)$ contribution in (\ref{LnaBI}) since it is
ambiguously defined, for example, terms like $F_{\mu \nu} F^{\nu
\rho} F_{\rho \sigma} F^{\sigma \nu}$ and $F_{\mu \nu} F^{\nu
\rho} F^{\sigma \nu} F_{\rho \sigma}$ which are equivalent from
the $abelian$ point of view are not so from the $nonabelian$ point
of view. The reason lies in the fact that the commutator of two
field strengths is not zero in the nonabelian case. So a
nonabelian generalization of the Born-Infeld lagrangian, in the
context of Superstring Theory is not an immediate task to
acheive.\\
A nonabelian calculation by means of a 4-point amplitude
calculation leads to the following expression
\cite{Tseytlin1,Gross1}:
\begin{eqnarray}
{\cal L}^{non-ab}_{\rm BI} & = & - \frac{1}{4} \mbox{tr}
\biggl(F_{\mu \nu} F^{\mu \nu} \biggr) + \frac{\pi^2}{2}
{\alpha'}^2 \mbox{tr} \biggl (\frac{1}{3} F_{\mu \nu} F^{\nu \rho}
F_{\rho \sigma} F^{\sigma \nu} + \frac{2}{3} F_{\mu \nu} F^{\nu
\rho} F^{\sigma \mu} F_{\rho \sigma} - \nonumber \\
&&-\frac{1}{6} F_{\mu \nu} F^{\mu \nu} F_{\rho \sigma} F^{\rho
\sigma} - \frac{1}{12} F_{\mu \nu} F_{\rho \sigma} F^{\mu \nu}
F^{\rho \sigma} \biggr)
 + {\cal O}({\alpha'}^3) \ .
\label{LnaBI1}
\end{eqnarray}
It is easy to see that the abelian limit of the ${\cal
O}({\alpha'}^2)$ terms in (\ref{LnaBI1}) agrees exactly with the
corresponding terms in (\ref{LBI1}).\\

\subsection{Introducing the symmetrized trace}

In \cite{Tseytlin2} it was seen that the abelian and the
nonabelian expressions of the Born-Infeld lagrangian could be
related, at least up to ${\cal O}({\alpha'}^2)$ terms, by introducing a symmetrized trace in the abelian expression (\ref{LBI1}),
\begin{eqnarray}
{\cal L}^{non-ab}_{\rm BI} & = &- \frac{1}{4} \mbox{str}
\biggl(F_{\mu \nu} F^{\mu \nu} \biggr)+ \frac{\pi^2}{2} {\alpha'}^2
\mbox{str} \biggl (F_{\mu \nu} F^{\nu \rho} F_{\rho \sigma} F^{\sigma \nu}
- \frac{1}{4} F_{\mu \nu} F^{\mu \nu} F_{\rho \sigma} F^{\rho \sigma} \biggr)
+ \nonumber \\
&& + {\cal O}({\alpha'}^3) \ . \label{LnaBI2}
\end{eqnarray}
The symmetrized trace, denoted by 'str', is defined as an average of
the trace of all possible permutations of matrices. The result in (\ref{LnaBI2}) is very nice in the sense that it looks like a democratic way of constructng the nonabelian lagrangian from the abelian one.\\
The complete proposal of \cite{Tseytlin2} for the nonabelian Born-Infeld lagrangian is simply
\begin{eqnarray}
{\cal L}^{non-ab}_{\rm BI} = (\mbox{constant}) \ \mbox{str} \biggl(
\sqrt{\mbox{det}(\eta_{\mu \nu} + 2 \pi \alpha' F_{\mu \nu})} \biggr)  \ .
\label{LBInastr}
\end{eqnarray}
This can be considered as a prescription for writing the $F^n$ terms of the lagrangian, at any order in $\alpha'$. Its abelian limit clearly agrees with the usual Born-Infeld lagrangian (\ref{LBI}).\\
So, the general structure of the low energy effective lagrangian in the open string sector of Type I theory may be written as
\begin{eqnarray}
{\cal L}_{\rm eff} = (\mbox{constant}) \ \mbox{str} \biggl(
\sqrt{\mbox{det}(\eta_{\mu \nu} + 2 \pi \alpha' F_{\mu \nu})} \biggr)
+ \biggl( \begin{array}{c}
          \mbox{covariant} \\
          \mbox{derivative} \\
          \mbox{terms}
          \end{array} \biggr) \ + (\mbox{fermions}) \ .
\label{LBInastr1}
\end{eqnarray}
Equation (\ref{LBInastr1}) may seem to be a quite simple and a very strong result and, although it is correct, it has a serious problem. Due to the $[D_{\mu}, D_{\nu}]F_{\alpha \beta} = -i g [F_{\mu \nu}, F_{\alpha \beta}]$ identity, the $F^n$ and the $D^{2p}F^{n-p}$ terms can be related, so the covariant derivative terms in (\ref{LBInastr1}) are as important as the $F^n$ ones. Therefore, the separation between $F^n$ terms (i.e. nonabelian Born-Infeld lagrangian) and covariant derivative ones is a purely artificial fact in the case of the nonabelian theory.\\
The conclusion is that the complete determination of the ${\cal L}_{\rm eff}$ lagrangian in (\ref{LBInastr1}) can only be obtained by perturbative theory in $\alpha'$.\\
The approach that we will follow in this work is the scattering amplitude one.

\section{Some details about the interactions}

\subsection{General formula for the string scattering amplitude}
\label{Symmetries}
\noindent The general formula for the (open string) massless boson amplitude (at tree level) is \cite{Schwarz1}
\begin{eqnarray}
{\cal A}^{(M)} = i \ (2 \pi)^{10} \delta(k_1 + k_2 + \ldots + k_M)
\cdot {\sum_{j_1, j_2, \ldots , j_M}}' \mbox{tr}(\lambda^{a_{j_1}}
\lambda^{a_{j_2}} \ldots \lambda^{a_{j_M}}) \ A(j_1, j_2, \ldots,
j_M) \ , \label{general-amplitude}
\end{eqnarray}
where $M$ is the number of bosons and the sum $\sum'$ in the indices $\{j_1, j_2, \ldots, j_M\}$
is done over $non-cyclic$ equivalent permutations of the group
$\{1, 2, \ldots, M\}$. The matrices $\lambda^{a_{j}}$ are in the
adjoint representation of the Lie group. $A(j_1,
j_2, \ldots, j_M)$ is the main object of study, called $subamplitude$. It corresponds to the
$M$-point amplitude of open superstrings which do not carry color
indices and which are placed in the ordering $\{j_1, j_2, \ldots,
j_M\}$ (modulo cyclic permutations). Using vertex operators, the RNS formalism leads to the following integral formula for $A(1,2,\ldots,M)$ \cite{Green1}, for $M \ge 3$:
\begin{eqnarray}
A(1, 2, \ldots,M) & = & 2 \frac{g^{M-2}}{(2 \alpha')^{7M/4+2}}
 (x_{M-1}-x_1)(x_{M}-x_1)\times \nonumber \\ && {}\times \int dx_2
 \ldots dx_{M-2} \int d \theta_1 \ldots d \theta_{M-2} \prod_{i>j}^M |
 x_i - x_j - \theta_i \theta_j |^{2 \alpha' k_i \cdot k_j}\times
\nonumber \\& &
{} \times \int d \phi_1 \ldots d \phi_M \ e^{f_M(\zeta, k, \theta,
 \phi)}\, ,
\label{N-amplitude}
\end{eqnarray}
where
\begin{equation}
f_M(\zeta, k, \theta, \phi) = \sum_{i \neq j}^M
\frac{(\theta_i-\theta_j) \phi_i (\zeta_i \cdot k_j) (2
\alpha')^{11/4} -1/2 \phi_i \phi_j (\zeta_i \cdot \zeta_j) (2
\alpha')^{9/2}}{x_i-x_j-\theta_i \theta_j}\,.
\label{fN}
\end{equation}
The $\theta_i$'s and the $\phi_i$'s in (\ref{N-amplitude}) and (\ref{fN}) are Grassmann variables, while the $x_i$'s are real variables such that $x_1 < x_2 < x_3 <\ldots < x_M$.
The $k_i$'s and the $\zeta_i$'s are the $i$-th string momentum and polarization, respectively.\\
Although not manifest, the subamplitude $A(1, 2, \ldots,M)$ in (\ref{N-amplitude}) has the following symmetries\cite{Schwarz1}:
\begin{enumerate}
\item Cyclicity:
$$A(1, 2, \ldots, M-1, M)=A(2, 3, \ldots, M, 1)= \ldots = A(M, 1, \ldots, M-2, M-1) \ .$$
\item On-shell gauge invariance:
$$\left. A(1, 2, \ldots, M) \right|_{\zeta_i=k_i} = 0, \ \ \mbox{for} \ \ i=1, \ldots, M \ .$$
\item World-sheet parity:
$$A(1, 2, \ldots, M-1, M) = (-1)^M A(M-1, M-2, \ldots, 1, M) \ .$$
\end{enumerate}
Formula (\ref{N-amplitude}), together with (\ref{general-amplitude}), contains $all$ the information to construct the low energy efective lagrangian, which has the form
\begin{eqnarray}
{\cal L}_{\rm eff} = \mbox{tr} \biggl\{ F^2 + {\alpha'}^2F^4 +
{\alpha'}^3 ( F^5 + D^2F^4 ) + {\alpha'}^4 ( F^6 + D^2F^5 + D^4F^4 ) + \ldots \ \biggr\} \ .
\label{generalLeff}
\end{eqnarray}
Formula (\ref{generalLeff}) already considers the fact that the string 3-point amplitude, $A(1,2,3)$, agrees completely with the corresponding Yang-Mills 3-point amplitude (i.e., it has no $\alpha'$ corrections \cite{Schwarz1}).\\

\subsection{Case of the 4-point amplitude}

An interesting (and very well known) application of formula (\ref{N-amplitude}) is the case of the 4-point subamplitude. It leads to
\begin{eqnarray}
A(1,2,3,4) & = & 8 \ g^2 \ {\alpha'}^2 \frac{\Gamma(-
\alpha' s) \Gamma(- \alpha' t)}{\Gamma(1- \alpha' s - \alpha' t)}
K(\zeta_1, k_1; \zeta_2, k_2; \zeta_3, k_3; \zeta_4, k_4) \ ,
\label{A1234}
\end{eqnarray}
where
\begin{eqnarray}
K(\zeta_1, k_1; \zeta_2, k_2; \zeta_3, k_3; \zeta_4, k_4) =
t_{(8)}^{\mu_1 \nu_1 \mu_2 \nu_2 \mu_3 \nu_3 \mu_4 \nu_4}
\zeta_{\mu_1}^1 k_{\nu_1}^1 \zeta_{\mu_2}^2 k_{\nu_2}^2
\zeta_{\mu_3}^3 k_{\nu_3}^3 \zeta_{\mu_4}^4 k_{\nu_4}^4
\label{kinematicfactor}
\end{eqnarray}
is a kinematic factor, $t_{(8)}$ being a known tensor
\cite{Schwarz1}. The $s$ and $t$ variables in (\ref{A1234}) are part
of the three Mandelstam variables. They may be written as \cite{DeRoo1}
\begin{eqnarray}
s = -k_1 \cdot k_2 - k_3 \cdot k_4, \ \ \, t = -k_1 \cdot k_4 - k_2 \cdot k_3 \ \ \ .
\label{Mandelstam}
\end{eqnarray}
The Gamma factor in (\ref{A1234}) has a completely known $\alpha'$
expansion, which begins like
\begin{eqnarray}
{\alpha'}^2 \frac{\Gamma(- \alpha' s) \Gamma(- \alpha' t)}
{\Gamma(1- \alpha' s - \alpha' t)} = \frac{1}{s t} - \frac{\
\pi^2}{6}{\alpha'}^2 + {\cal O}({\alpha'}^3)\,. \label{gammas}
\end{eqnarray}
Using (\ref{Mandelstam}) and the known symmetries of the $t_{(8)}$ tensor \cite{Schwarz1}, the expression for $A(1,2,3,4)$ in (\ref{A1234}) has all three symmetries (cyclic invariance, on-shell gauge invariance and world-sheet parity) \underline{manifest}.\\
The general method of finding the string corrections to the Yang-Mills lagrangian, at a given $\alpha'$ order, consists in writing all the possible terms with unknown coefficients. Consider, for example, the ${\cal O}({\alpha'}^2)$ corrections. Up to that $\alpha'$ order the effective lagrangian looks like \cite{Tseytlin1}
\begin{eqnarray}
{\cal L}_{\rm eff} & = & - \frac{1}{4} \mbox{tr}
\biggl(F_{\mu \nu} F^{\mu \nu} \biggr) + \frac{\pi^2}{2}
{\alpha'}^2 \mbox{tr} \biggl (c_1 F_{\mu \nu} F^{\nu \rho}
F_{\rho \sigma} F^{\sigma \nu} + c_2 F_{\mu \nu} F^{\nu
\rho} F^{\sigma \mu} F_{\rho \sigma} + \nonumber \\
&&+ c_3 F_{\mu \nu} F^{\mu \nu} F_{\rho \sigma} F^{\rho
\sigma} + c_4 F_{\mu \nu} F_{\rho \sigma} F^{\mu \nu}
F^{\rho \sigma} \biggr) \ .
\label{Leff2}
\end{eqnarray}
This expression is the final result, after having considered also derivative terms and used the Bianchi identity, integration by parts and the $[D, D]F = [F,F]$ relations. In (\ref{Leff2}), we have also omitted terms which vanish on-shell (i.e., terms which contain $D_{\mu} F^{\mu \nu}$). \\
Then, the tree level 4-point amplitude is calculated using (\ref{Leff2}) and compared with the corresponding expression in (\ref{A1234}) up to ${\cal O}({\alpha'}^2)$ order. This comparison determines the unknowns $c_1$, $c_2$, $c_3$ and $c_4$, leading to the expression in (\ref{LnaBI1}) \cite{Tseytlin1}).\\
Although this procedure can be applied to compute the correction terms (sensible to a 4-point amplitude) up to any $\alpha'$ order, it gets longer and harder as the $\alpha'$ order grows.\\
Based on the idea of \cite{DeRoo1}, in \cite{Chandia1} was done the explicit construction of all $D^{2n} F^4$ ($n=0, 1, 2, \ldots $) in the effective lagrangian, arriving to
\begin{eqnarray}
{\cal L}_{D^{2n}F^4}  & = & - \frac{1}{8} g^2 {\alpha'}^2 \
 \int \int \int \int \
\left\{ \frac{}{}\prod_{j=1}^{4} d^{10} x_j \ \delta^{(10)}(x-x_j)
\frac{}{} \right\} \times \nonumber \\
&& \times f_{\rm sym} \left( \frac{}{} \frac{ (D_{1} + D_{2})^2 +
(D_{3} + D_{4})^2}{2} ,
\frac{(D_{1} + D_{4})^2 + (D_{2} + D_{3})^2}{2} \frac{}{} \right) \times \nonumber \\
&& \times t_{(8)}^{\mu_1 \nu_1 \mu_2 \nu_2 \mu_3 \nu_3 \mu_4 \nu_4}
\mbox{tr} \biggl( F_{\mu_1 \nu_1}(x_1)F_{\mu_2 \nu_2}(x_2)F_{\mu_3
\nu_3}(x_3)F_{\mu_4 \nu_4}(x_4) \biggr) \ ,
\label{Seff-nonabelian}
\end{eqnarray}
where the function $f$ is given by
\begin{eqnarray}
f(s, t) & = & \frac{\Gamma(- \alpha' s)\Gamma(- \alpha' t)}
{\Gamma(1 - \alpha' s - \alpha' t)} - \frac{1}{{\alpha'}^2 st} \ .
\label{f}
\end{eqnarray}
(See \cite{Chandia1} for more details about the relation between functions $f_{\rm sym}(s,t)$ and $f(s,t)$.)

\subsection{Case of the 5-point amplitude}

\noindent At this point is where the search for $higher$ N-point amplitudes begins. In the case of the 5-point subamplitude formula (\ref{N-amplitude}) becomes
\begin{eqnarray}
A(1, 2, 3, 4, 5) & = & 2 \frac{g^3}{(2 \alpha')^{43/4}} (x_4-x_1)
(x_5-x_1)\times
\nonumber \\&&{}\times  \int_{x_1}^{x_4} d x_3 \int_{x_1}^{x_3} d x_2
 \int
d \theta_1 d \theta_2 d \theta_3 \prod_{i>j}^5 | x_i - x_j - \theta_i
\theta_j |^{2 \alpha' k_i \cdot k_j}\times
\nonumber \\&&{}\times  \int d \phi_1 d \phi_2 d \phi_3 d \phi_4 d
 \phi_5
e^{f_5(\zeta, k, \theta, \phi)}\,,
\label{A12345}
\end{eqnarray}
where $\theta_4= \theta_5 =0$.\\
Once the Grassmann integration has been done, and after some lengthly algebra, it can be written as \cite{Brandt1}
\begin{eqnarray}
A(1,2,3,4,5) & = & 2 \ g^3 \ (2 \ \alpha')^2 \biggl\{ L_3 (\zeta_1 \cdot \zeta_2)(\zeta_3 \cdot \zeta_4)(\zeta_5 \cdot k_2)(k_1 \cdot k_3)
+ \biggl( \mbox{44} \ (\zeta \cdot \zeta)^2 (\zeta \cdot k)(k \cdot k) \ \mbox{terms} \biggr) \nonumber \\
&&\phantom{2 \ g^3 \ (2 \ \alpha')^2 \biggl\{ } + K_2 (\zeta_1 \cdot \zeta_4)(\zeta_5 \cdot k_2)(\zeta_2 \cdot k_1)(\zeta_3 \cdot k_4)
+ \biggl( \mbox{99} \ (\zeta \cdot \zeta) (\zeta \cdot k)^3 \ \mbox{terms} \biggr) \biggr\} .
\nonumber \\
\label{A12345-1}
\end{eqnarray}
(See eq. (5.29) of \cite{Brandt1} for the complete detailed formula.) $L_3$ and $K_2$ are momentum dependent factors (which also depend on $\alpha'$) given by double integrals:
\begin{eqnarray}
\biggl\{ \begin{array}{c}
         L_3 \\
         K_2
         \end{array} \biggr\} =
\int_0^1 dx_3 \int _0^{x_3} dx_2 \ x_2^{2 \alpha' \alpha_{12}} (1-x_2)^{2 \alpha' \alpha_{24}} x_3^{2 \alpha' \alpha_{13}} (1-x_3)^{2 \alpha' \alpha_{34}} (x_3-x_2)^{2 \alpha' \alpha_{23}}
\biggl\{ \begin{array}{c}
         \frac{1}{x_2 x_3 (1-x_3)} \\
         \frac{1}{x_2 (1-x_3)}
         \end{array} \biggr\} \ , \nonumber\\
\label{L3K2}
\end{eqnarray}
where $\alpha_{ij}=k_i \cdot k_j$. They can be calculated in terms of Beta and hypergeometric functions. Independently of the method used to calculate them, the first terms of their $\alpha'$ expansion can be obtained. For example,
\begin{eqnarray}
K_2 & = & \frac{1}{(2 \alpha')^2} \left\{ \frac{1}{\alpha_{12} \
\alpha_{34}} \right\} - \frac{ \ \pi^2}{6} \left\{\frac{
\alpha_{51} \ \alpha_{12} - \alpha_{12} \ \alpha_{34} +
\alpha_{34} \ \alpha_{45}}{\alpha_{12} \ \alpha_{34}} \right\} +
\nonumber
\\&+& \zeta(3) \ (2 \alpha') \left\{  \frac{\alpha_{12}^2 \ \alpha_{51}
 - \alpha_{34}^2 \ \alpha_{12} +
\alpha_{45}^2 \ \alpha_{34} + \alpha_{51}^2 \ \alpha_{12} -
\alpha_{12}^2 \ \alpha_{34} + \alpha_{34}^2 \ \alpha_{45} - 2
\alpha_{12} \ \alpha_{23} \ \alpha_{34} }{\alpha_{12} \
\alpha_{34}} \right\} \
+ \nonumber\\
&+& {\cal O}((2\alpha')^2) \,. \label{K2}
\label{K2}
\end{eqnarray}
Formula (\ref{A12345-1}) was used in \cite{Brandt1} to distinguish from three non equivalent versions of the ${\alpha'}^3F^5$ terms of (\ref{generalLeff}) \cite{Kitazawa1, Refolli1, Koerber1}, obtaining complete agreement with the one in \cite{Koerber1}.\\
In \cite{Brandt1}, formula (\ref{A12345-1}) was written in terms of 8 $K_i$'s and 8 $L_i$'s (plus cyclic permutations of the terms in the amplitude).  It was seen that some linear relations, coming from integration by parts technique, existed between the $K_i$'s and between the $L_i$'s.\\
Afterwards, in \cite{Machado1} it was seen that further linear relations existed between those $\alpha'$ dependent factors. The new relations, independent from the ones obtained in \cite{Brandt1}, were due to the partial fraction technique. So at the end, for the 16 $\alpha'$ dependent factors there were found 7 independent integration by parts relations and 7 independent partial fraction relations. Solving this linear system it allowed to write $all$ $K_i$'s and $L_i$'s in terms of only 2 of them. This was summarized and exploited in \cite{Barreiro1}, where the final expression for the 5-point subamplitude was written as
\begin{eqnarray}
A(1,2,3,4,5) = T \cdot A_{YM}(1,2,3,4,5) + (2  \alpha')^2 K_3
\cdot A_{F^4}(1,2,3,4,5) \ .
\label{A12345final}
\end{eqnarray}
In this formula $A_{YM}(1,2,3,4,5)$ and $A_{F^4}(1,2,3,4,5)$ are the 5-point subamplitudes coming from the Yang-Mills and the known $F^4$ terms in (\ref{LnaBI}), while $T$ and $K_3$ are $\alpha'$ dependent factors which have a known $\alpha'$ expansion which go like
\begin{eqnarray}
T = 1 + {\cal O}({\alpha'^3}) \ , \ \ \
(2 \alpha')^2 K_3 = \frac{\pi^2}{6} (2 \alpha')^2 + {\cal O}({\alpha'^3}) \ .
\label{TK3}
\end{eqnarray}
Formula (\ref{A12345final}), the same as formula (\ref{A1234}), has the nice property that the cyclic, the on-shell gauge invariance and the world-sheet parity symmetries are \underline{manifest} (see \cite{Barreiro1} for further details). We have also tested the factorization properties of the poles \cite{Barreiro1}. \\

\noindent \underline{\bf Benefits from having a closed formula for A(1,2,3,4,5):}

\subsubsection{5-point amplitudes involving fermions are immediate}
\noindent By this we mean that there is no need to compute the 3-boson/2-fermion and the 1-boson/4-fermion amplitudes right from the beginning, in the RNS formalism. The explanation is the following. Up to now, the supersymmetric low energy effective lagrangian is $completely$ known up to ${\cal O}({\alpha'}^2)$ terms. It has the form
\begin{eqnarray}
{\cal L}_{\rm eff} = {\cal L}_{\rm SYM} + {\alpha'}^2 {\cal L}_2
+ {\cal O}({\alpha'}^3) \ ,
\label{Leff3}
\end{eqnarray}
where
\begin{eqnarray}
{\cal L}_{\rm SYM} = \mbox{tr} \biggl( F^2 + i \bar{\psi} \gamma D \psi \biggr)
\label{LSYM}
\end{eqnarray}
is the D=10 Super Yang-Mills lagrangian and
\begin{eqnarray}
{\cal L}_2 = \mbox{tr} \biggl( F^4 + D (\bar{\psi} \gamma \psi) F^2 +
D^2 (\bar{\psi} \gamma \psi)^2 + F (\bar{\psi} \gamma \psi)^4   \biggr)
\label{L2}
\end{eqnarray}
is the order ${\alpha'}^2$ string correction to ${\cal L}_{\rm SYM}$. ${\cal L}_2$ has been determined completely in \cite{Cederwall1}.\\
In this sense, due to the structure of the $\alpha'$ expansion of $T$ and $(2 \alpha')^2 K_3$ in (\ref{TK3}), we could rewrite eq. (\ref{A12345final}) in a slightly different (but equivalent) notation:
\begin{eqnarray}
A^{5b}(1,2,3,4,5) = T \cdot A^{5b}_{SYM}(1,2,3,4,5) + (2  \alpha')^2 K_3
\cdot A^{5b}_{{\cal L}_2}(1,2,3,4,5) \ ,
\label{A12345-5b}
\end{eqnarray}
where $A^{5b}_{SYM}(1,2,3,4,5)$ and $A^{5b}_{{\cal L}_2}(1,2,3,4,5)$ denote the 5-boson subamplitude coming from ${\cal L}_{\rm SYM}$ and ${\cal L}_2$, respectively.\\
Now we write down our `immediate' expression for the 3-boson/2-fermion and the 1-boson/4-fermion subamplitudes. Our ansatz, based on the structure of the $\alpha'$ expansion of $T$ and $(2 \alpha')^2 K_3$ in (\ref{TK3}), is the following, respectively \cite{Barreiro2}:
\begin{eqnarray}
\label{A12345-3b-2f}
A^{3b/2f}(1,2,3,4,5) = T \cdot A^{3b/2f}_{SYM}(1,2,3,4,5) + (2  \alpha')^2 K_3
\cdot A^{3b/2f}_{{\cal L}_2}(1,2,3,4,5) \ , \\
\label{A12345-1b-4f}
A^{1b/4f}(1,2,3,4,5) = T \cdot A^{1b/4f}_{SYM}(1,2,3,4,5) + (2  \alpha')^2 K_3
\cdot A^{1b/4f}_{{\cal L}_2}(1,2,3,4,5) \ .
\end{eqnarray}
As an immediate test of (\ref{A12345-3b-2f}) and (\ref{A12345-1b-4f}) we see that, by construction they reproduce the 3-boson/2-fermion and the 1-boson/4-fermion subamplitudes of the low energy effective lagrangian in (\ref{Leff3}). On the other side, and this guarantees that (\ref{A12345-3b-2f}) and (\ref{A12345-1b-4f}) are correct to any order in $\alpha'$, these formulas, together with (\ref{A12345-5b}), satisfy by construction the supersymmetry requirement: the summed variation of $A^{5b}(1,2,3,4,5)$, $A^{3b/2f}(1,2,3,4,5)$  and $A^{1b/4f}(1,2,3,4,5)$ under the supersymmetry transformations \cite{Green1},
\begin{eqnarray}
\delta A^a_{\mu} = \frac{i}{2} \bar{\epsilon} \gamma_{\mu} \psi^a \ , \\
\delta \psi^a = -\frac{1}{4} F^a_{\mu \nu} \gamma^{\mu \nu} \epsilon \ , \\
\delta \bar{\psi}^a = -\frac{1}{4} \bar{\epsilon} \gamma^{\mu \nu} F^a_{\mu \nu} \ ,
\label{susytransf}
\end{eqnarray}
is zero, after using the on-shell and the physical state conditions, together with momentum conservation.\\
Formula (\ref{A12345-1b-4f}) is being used in \cite{Barreiro2} to determine the ${\alpha'}^3 D^2 F (\bar{\psi} \gamma \psi)^2$ terms, which are unknown at the present moment.

\subsubsection{Determination of the ${\alpha'}^{n+3} D^{2n} F^5$ terms}

\noindent In (\ref{Seff-nonabelian}) it was seen that it was possible to explicitly construct all the effective lagrangian terms which are sensible to the 4-point
amplitude.\\
We will now see that that using the compact formula (\ref{A12345final}) we have been able to determine all the ${\alpha'}^{n+3} D^{2n} F^5$ terms in (\ref{generalLeff}). For this purpose we compute first the corresponding scattering amplitude as
\begin{eqnarray}
A_{D^{2n}F^5}(1,2,3,4,5) = A(1,2,3,4,5) - A_{YM}(1,2,3,4,5) -
A_{D^{2n}F^4}(1,2,3,4,5) \ .
\label{AD2nF5}
\end{eqnarray}
It is quite remarkable that the resulting expression has no poles, as it should happen (see \cite{Barreiro1} for further details). With the simplified expression for $A_{D^{2n}F^5}(1,2,3,4,5)$ we find the corresponding lagrangian terms to be \cite{Barreiro1}:
\begin{eqnarray}
{\cal L}_{D^{2n}F^5}  =  \ i \ g^3 \ \int \int \int \int \int
\left\{ \frac{}{}\prod_{j=1}^{5} d^{10} x_j \ \delta^{(10)}(x-x_j)
\frac{}{} \right\} \times \hspace{5.5cm} \nonumber \\
 \times  \ \biggl[ \ \biggl\{ \frac{1}{32} H^{(1)} ( - D_{2} \cdot
D_{3}, - D_{3} \cdot D_{4} \ , - D_{4} \cdot D_{5} ) \
t_{(10)}^{\mu_1 \nu_1 \mu_2 \nu_2
\mu_3 \nu_3 \mu_4 \nu_4 \mu_5 \nu_5} + \nonumber \\
+ \frac{1}{16} P^{(1)} ( - D_{1} \cdot D_{2}, - D_{2} \cdot D_{3},
- D_{3} \cdot D_{4} \ , - D_{4} \cdot D_{5}, - D_{5} \cdot D_{1} ) \
{(\eta \cdot t_{(8)})_1}^{\mu_1 \nu_1 \mu_2 \nu_2
\mu_3 \nu_3 \mu_4 \nu_4 \mu_5 \nu_5} \biggr\} \times \nonumber \\
\times \mbox{tr}\biggl( F_{\mu_1 \nu_1}(x_1)F_{\mu_2 \nu_2}(x_2)F_{\mu_3
\nu_3}(x_3)F_{\mu_4 \nu_4}(x_4)F_{\mu_5 \nu_5}(x_5) \biggr) \ \ - \nonumber
\\
- \ \ U^{(1)} ( - D_{1} \cdot D_{2}, - D_{2} \cdot D_{3}, - D_{4}
\cdot
D_{5}, - D_{5} \cdot D_{1} ) \times \nonumber \\
\times \biggl\{\frac{1}{64} t_{(10)}^{\mu_1 \nu_1 \mu_2 \nu_2
\mu_3 \nu_3 \mu_4 \nu_4 \mu_5 \nu_5} \mbox{tr}\biggl( F_{\mu_1
\nu_1}(x_1)F_{\mu_2 \nu_2}(x_2)D^{\alpha}F_{\mu_3
\nu_3}(x_3)D_{\alpha} F_{\mu_4 \nu_4}(x_4)F_{\mu_5 \nu_5}(x_5)
\biggr) + \nonumber \\
+ \frac{1}{16} t_{(8)}^{\mu_4 \nu_4 \mu_5 \nu_5 \mu_1 \nu_1 \mu_2
\nu_2} \mbox{tr}\biggl( D^{\mu_3}F_{\mu_1 \nu_1}(x_1)D^{\nu_3}F_{\mu_2
\nu_2}(x_2)F_{\mu_3 \nu_3}(x_3)F_{\mu_4 \nu_4}(x_4)F_{\mu_5
\nu_5}(x_5) \biggr)+ \nonumber \\
+ \frac{1}{16} t_{(8)}^{\mu_3 \nu_3 \mu_4 \nu_4 \mu_5 \nu_5 \mu_1
\nu_1} \mbox{tr}\biggl( D^{\mu_2}F_{\mu_1 \nu_1}(x_1)F_{\mu_2
\nu_2}(x_2)D^{\nu_2}F_{\mu_3 \nu_3}(x_3)F_{\mu_4 \nu_4}(x_4)F_{\mu_5
\nu_5}(x_5) \biggr)- \nonumber \\
- \frac{1}{16} t_{(8)}^{\mu_1 \nu_1 \mu_2 \nu_2 \mu_3 \nu_3 \mu_4
\nu_4} \mbox{tr}\biggl( D^{\mu_5}F_{\mu_1 \nu_1}(x_1)F_{\mu_2
\nu_2}(x_2)F_{\mu_3 \nu_3}(x_3)D^{\nu_5}F_{\mu_4 \nu_4}(x_4)F_{\mu_5
\nu_5}(x_5) \biggr)- \nonumber \\
- \frac{1}{16} t_{(8)}^{\mu_5 \nu_5 \mu_1 \nu_1 \mu_2 \nu_2
\mu_3 \nu_3} \mbox{tr}\biggl( D^{\mu_4}F_{\mu_1 \nu_1}(x_1)F_{\mu_2
\nu_2}(x_2)F_{\mu_3 \nu_3}(x_3)F_{\mu_4
\nu_4}(x_4)D^{\nu_4}F_{\mu_5 \nu_5}(x_5) \biggr) \ \biggr\} \ \
- \nonumber \\
- \ \ \frac{1}{8} W^{(1)} ( - D_{1} \cdot D_{2}, - D_{2} \cdot
D_{3}, - D_{3} \cdot D_{4} \ , - D_{4} \cdot D_{5}, - D_{5} \cdot
D_{1} )
\times \nonumber \\
\times t_{(8)}^{\mu_2 \nu_2 \mu_3 \nu_3 \mu_4 \nu_4 \mu_5 \nu_5}
\mbox{tr}\biggl( F_{\mu_1 \nu_1}(x_1)D^{\mu_1}F_{\mu_2
\nu_2}(x_2)F_{\mu_3 \nu_3}(x_3)F_{\mu_4
\nu_4}(x_4)D^{\nu_1}F_{\mu_5 \nu_5}(x_5) \biggr) \ \ - \nonumber \\
- \ \
\frac{1}{8} Z^{(1)} ( - D_{1} \cdot D_{2}, - D_{2} \cdot D_{3}, -
D_{3} \cdot D_{4} \ , - D_{4} \cdot D_{5}, - D_{5} \cdot D_{1} )
\times \nonumber \\
\times t_{(8)}^{\mu_2 \nu_2 \mu_3 \nu_3 \mu_4 \nu_4 \mu_5 \nu_5}
\biggl\{ \mbox{tr}\biggl( F_{\mu_1 \nu_1}(x_1)F_{\mu_2
\nu_2}(x_2)D^{\mu_1}F_{\mu_3 \nu_3}(x_3)D^{\nu_1}F_{\mu_4
\nu_4}(x_4)F_{\mu_5 \nu_5}(x_5) \biggr)
- \nonumber \\
\hphantom{\times t_{(8)}^{\mu_2 \nu_2 \mu_3 \nu_3 \mu_4 \nu_4
\mu_5 \nu_5} } - \mbox{tr}\biggl( F_{\mu_1 \nu_1}(x_1)D^{\mu_1}F_{\mu_2
\nu_2}(x_2)F_{\mu_3 \nu_3}(x_3)F_{\mu_4 \nu_4}(x_4)D^{\nu_1}F_{\mu_5
\nu_5}(x_5) \biggr) \biggr\} \ \ + \nonumber \\
+ \ \ \frac{1}{160} \Delta( - D_{1} \cdot D_{2}, - D_{2} \cdot
D_{3}, - D_{3}
\cdot D_{4} \ , - D_{4} \cdot D_{5}, - D_{5} \cdot D_{1} ) \times \nonumber \\
\times \biggl\{ t_{(10)}^{\mu_1 \nu_1 \mu_2 \nu_2 \mu_3 \nu_3
\mu_4 \nu_4 \mu_5 \nu_5}  \mbox{tr}\biggl( D^{\alpha} D^{\beta} F_{\mu_1
\nu_1}(x_1)D_{\alpha}F_{\mu_2 \nu_2}(x_2)F_{\mu_3
\nu_3}(x_3)F_{\mu_4
\nu_4}(x_4)D_{\beta}F_{\mu_5 \nu_5}(x_5) \biggr) + \nonumber \\
+ 4 {(\eta \cdot t_{(8)})_1}^{\mu_1 \nu_1 \mu_2 \nu_2 \mu_3 \nu_3
\mu_4 \nu_4 \mu_5 \nu_5}  \times \nonumber \\
\hphantom{ + 4 }  \times \mbox{tr}\biggl( F_{\mu_1
\nu_1}(x_1)D_{\alpha}F_{\mu_2 \nu_2}(x_2)D^{\alpha} F_{\mu_3
\nu_3}(x_3)D_{\beta} F_{\mu_4 \nu_4}(x_4)D^{\beta}F_{\mu_5
\nu_5}(x_5) \biggr) \ \biggr\} \ \ \ \biggr] \ , \nonumber \\
\label{LD2nF5final}
\end{eqnarray}
where $H^{(1)}$, $P^{(1)}$, $U^{(1)}$, $W^{(1)}$, $Z^{(1)}$ and $\Delta$ have known expressions (and therefore known $\alpha'$ expansions) in terms of the Gamma factor, $T$ and $K_3$. \\
In (\ref{LD2nF5final}) we see that, besides the known $t_{(8)}$ tensor, a new $t_{(10)}$ tensor has arisen \cite{Barreiro1}.  Formula (\ref{LD2nF5final}) has been tested to reproduce the already known ${\alpha'}^3 F^5$ terms and gives the explicit construction of all covariant derivative terms containing 5 $F$'s (which are sensible to the 5-point amplitude).

\section{Towards a closed formula for N-point (tree level) amplitudes in Open Superstring Theory}
\noindent The method of finding a basis of $\alpha'$ dependent factors that allows to write the scattering amplitude in a short form (see eqs. (\ref{A1234}) and (\ref{A12345final})), by using `integration by parts' and `partial fractions' techniques, can be used for any N-point amplitude ($N > 3$) and will lead to an expression of the form
\begin{eqnarray}
A(1, 2, \ldots, N) = F_1(\alpha_{ij}; \alpha') K_1(\zeta,k) + \ldots +
F_{m_n}(\alpha_{ij}; \alpha') K_{m_n}(\zeta,k) \ .
\label{N-point}
\end{eqnarray}
Here, the $F_p(\alpha_{ij}; \alpha')$'s are the $\alpha'$ dependent factors and the
$K_p(\zeta,k)$'s are the kinematical expressions. By now, the completely known cases are only the $N=4$ and the $N=5$ ones.\\
The ambitious program would consist then in:
\begin{enumerate}
\item Finding how many terms are there in formula (\ref{N-point}): $m_n = ?$
\item Finding the specific formulas for:
\begin{itemize}
\item The kinematical expressions $K_p(\zeta,k)$'s, in such a way that the tensors
$t_{(8)}$, $t_{(10)}$, $\ldots$, $t_{(m_n)}$ can be determined.
\item The $\alpha'$ factors $F_p(\alpha_{ij}; \alpha')$'s and its $\alpha'$ expansions.
\end{itemize}
\end{enumerate}
This is still an open problem. In order to get an insight it would be good to consider the case of the 6-point amplitude, but before that we will make an important comment about how the symmetries of the scattering amplitude restrict its kinematical expression.\\
\subsection{Implementing symmetries in the scattering amplitude}
\label{Symmetries2}
\noindent In subsection \ref{Symmetries} it was seen that the N-point subamplitude $A(1,2, \ldots, N)$ satisfies cyclicity, on-shell gauge invariance and world-sheet parity. We have verified, in the case of $N=4$ and $N=5$ that, after doing all Grassmann integrations in (\ref{N-amplitude}) (like in (\ref{A12345-1}), in the case of $N=5$) and demanding the 3 symmetries to be satisfied, then a set of linear relations between the $\alpha'$ factors is found. This system of relations happens to be linearly equivalent to the one obtained when using `integration by parts' and `partial fractions' techniques. We have verified that this works correctly also in the case of bosonic string amplitudes (which have an expression similar to the one in (\ref{N-amplitude}), but with no Grassmann variables), for $N=4$ and $N=5$.\\
The deeper mean of all this is that, at least up to $N=5$, the symmetries are enough to fix the kinematics that governs the scattering amplitude, at any order in $\alpha'$. But, unfortunately, as we will see in the next subsection, this is no longer true already when $N=6$.\\
\subsection{Case of the 6-point amplitude}
\noindent After doing the Grassmann integration in (\ref{N-amplitude}) in the case of $N=6$ we arrive to an expression of the following type\cite{Barreiro3}:
\begin{eqnarray}
A(1,2,3,4,5,6) & = & (2 \ \alpha')^2 \ g^4 \ \biggl\{ \ \biggl((2 \alpha') I_6 (\zeta_4 \cdot \zeta_5)(\zeta_6 \cdot k_2)(\zeta_1 \cdot k_5)(\zeta_2 \cdot k_4)(\zeta_3 \cdot k_4) + \nonumber \\
&&\phantom{(2 \ \alpha')^3 \ g^4 \  \biggl\{ } + ( \mbox{other} \ (\zeta \cdot \zeta) (\zeta \cdot k)^4 \ \mbox{terms} ) \biggr) \ + \nonumber \\
&&\phantom{(2 \ \alpha')^3 \ g^4 \  \biggl\{ } + \ \biggl((1-2 \alpha' k_3 \cdot k_4) I_{45} (\zeta_1 \cdot \zeta_2)(\zeta_3 \cdot \zeta_4)(\zeta_5 \cdot k_1)(\zeta_6 \cdot k_2) + \nonumber \\
&&\phantom{(2 \ \alpha')^3 \ g^4 \  \biggl\{ } + ( \mbox{other} \ (\zeta \cdot \zeta)^2 (\zeta \cdot k)^2 \ \mbox{terms} ) \biggr) \ + \nonumber \\
&&\phantom{(2 \ \alpha')^3 \ g^4 \  \biggl\{ } + \ \biggl((2 \alpha') I_{109} (\zeta_2 \cdot \zeta_3)(\zeta_4 \cdot \zeta_5)(\zeta_6 \cdot \zeta_1)(k_1 \cdot k_2)(k_3 \cdot k_4) + \nonumber \\
&&\phantom{(2 \ \alpha')^3 \ g^4 \  \biggl\{ } + ( \mbox{other} \ (\zeta \cdot \zeta)^3 (k \cdot k)^2 \ \mbox{terms} ) \biggr) \ \biggr\} \ ,
\label{A12346-1}
\end{eqnarray}
where $I_6$, $I_{45}$ and $I_{109}$ are $\alpha'$ dependent factors given by triple integrals:
\begin{eqnarray}
\biggl\{ \begin{array}{c}
         I_6 \\
         I_{45} \\
         I_{109}
         \end{array} \biggr\} & = &
\int_0^1 dx_4 \int _0^{x_4} dx_3 \ \int _0^{x_3} dx_2 \ x_2^{2 \alpha' \alpha_{12}} (1-x_2)^{2 \alpha' \alpha_{25}} x_3^{2 \alpha' \alpha_{13}} (1-x_3)^{2 \alpha' \alpha_{35}}
x_4^{2 \alpha' \alpha_{14}} (1-x_4)^{2 \alpha' \alpha_{45}} \nonumber \\
&& \times (x_3-x_2)^{2 \alpha' \alpha_{23}} (x_4-x_2)^{2 \alpha' \alpha_{24}} (x_4-x_3)^{2 \alpha' \alpha_{34}} \ \times
\biggl\{ \begin{array}{c}
         \frac{1}{(1-x_4) (x_4-x_2)(x_4-x_3)} \\
         \frac{1}{x_2 (x_4-x_3)^2} \\
         \frac{1}{x_2 (1-x_4) (x_3-x_2)(x_4-x_3)}
         \end{array} \biggr\} \ . \nonumber\\
\label{I6I45I109}
\end{eqnarray}
Formula (\ref{A12346-1}) contains at all 237 different $\alpha'$ dependent factors $I_j$.\\
Demanding the symmetries to be satisfied, as explained in subsection \ref{Symmetries2}, we obtain a set of linear relations which allows us to write the $\alpha'$ dependent factors in terms of 15 of them.\\
On the past year an interesting preprint appeared with the calculation of the 6-point amplitude\cite{Stieberger1}. The authors of it did not find the relations between the $\alpha'$ dependent factors by means of integration by parts neither by partial fractions techniques (although they saw that some of their relations matched with the ones that come from these techniques). They demanded another symmetry to be satisfied by the integral expression of $A(1, \ldots, M)$: the superdiffeomorphism invariance on the string world-sheet. In fact they work with an integral expression for the amplitude which is not exactly the same as we wrote in (\ref{N-amplitude}), where we have already admitted $x_1$, $x_{M-1}$ and $x_M$ to be fixed (and not integrated) and also where we had fixed $\theta_{M-1}=\theta_M=0$. Their important result consists in the fact that they find a basis containing 6 $\alpha'$ dependent factors (instead of the 15 dimensional basis that we found). The test that supports their result consists in the fact that the linear system of equations that they find contains a lot more equations than unknowns (and still it is consistent).\\
Although all the authors of \cite{Stieberger1} give the first terms of the $\alpha'$ expansions of the 6 factors, they do not make any confirmaton between their expression and the 6-point amplitude that comes from the $D=10$ low energy effective lagrangian (\ref{generalLeff}). It would have been nice if they had checked the known terms up to ${\cal O}({\alpha'}^3)$ order (that have already been checked by S-matrix calculations and other methods) and, moreover, they had confirmed the ${\cal O}({\alpha'}^4)$ terms obtained in \cite{Koerber2}.\\
Anyway, the fact of their basis being 6-dimensional, motivated us to find the linear relations between the $\alpha'$ dependent factors directly by considering the integration by parts and the partial fractions techniques. The result of our computations agreed with their result: the basis is 6-dimensional. So, besides finding agreement with the dimension of the basis of the $I_j$'s space, we conclude that demanding cyclicity, on-shell gauge invariance and world-sheet parity in the scattering amplitude, it does no longer determine the kinematics completely. In the same way, it is not guaranteed that the method proposed in \cite{Stieberger1} will be equivalent to the integration by parts and partial fractions technique when $N>6$.\\
Up to this moment we have not obtained an explicit closed form for $A(1,2,3,4,5,6)$ since the expressions for the $I_j$'s in terms of the ones in the basis are extremely huge.\\

\section{Final remarks and conclusions}
\noindent We finish summarizing the main points of this talk:
\begin{itemize}
\item There does exist a method to \underline{explicitly} compute tree level scattering amplitudes in Open Superstring Theory, beyond 4-point calculations. The method is based on the `integration by parts' and `partial fraction' techniques of Integral Calculus, for the $\alpha'$ dependent factors that show up in the subamplitude. Any other method, which demands any kind of symmetry present in the scattering amplitudes, should lead to equivalent linear relations for those factors.
\item The method has been successfully applied to compute all massless 5-point amplitudes in Open Superstring Theory (5 boson, 3-boson/2-fermion and 1-boson/4-fermion).
\item The N-point case is still an open problem.
\item In order to look for some generalization, in the N-point case, it would be good to have a closed formula for the 6-point amplitude which had been tested to reproduce the effective lagrangian terms up to ${\cal O}({\alpha'}^4)$ order. There is some work in progress in this direction.
\item The kind of results presented in this talk are of importance in:
\begin{enumerate}
\item Determining the complete low energy effective lagrangian (at least in the open superstring sector) in Type I theory.
\item Loop amplitudes: it is quite probable that the same kinematic expressions that already appear at tree level also show up in higher loop calculations.
\item The low energy effective lagrangian of the Type II theories, since closed string amplitudes can be directly obtained from the open ones (by means of the KLT relations \cite{Kawai1}).
\end{enumerate}
\end{itemize}

\section*{Acknowledgements}

\noindent R. M. would like to acknowledge useful conversations with N.
Berkovits and is grateful to the organizers of the Fifth International Conference on Mathematical Methods in Physics - IC2006 for having invited him to give this seminar.


\begin{thebibliography}{99}
\bibitem{Fradkin1}
E.S. Fradkin and A.A. Tseytlin, \emph{Nonlinear electrodynamics
from quantized strings}, \emph{Phys. Lett.}{\bf B 163}(1985), 123.

\bibitem{Tseytlin1}
A.A. Tseytlin, \emph{Vector field effective action in the open
  superstring theory}, \emph{Nucl. Phys.} {\bf B 276}(1986), 391; Erratum-ibid:
  \emph{Nucl. Phys.}{\bf B 291}(1987),876.

\bibitem{Gross1}
D.J. Gross and E.~Witten, \emph{Superstring modifications of
  Einstein's equations}, \emph{Nucl. Phys.} {\bf B 277}(1986),1.

\bibitem{Tseytlin2}
A.A. Tseytlin, \emph{On non-abelian generalization of Born-Infeld
  action in string theory}, \emph{Nucl. Phys.} {\bf B 501}(1997), 41 [{\tt hep-th/9701125}].

\bibitem{Schwarz1}
J. H. Schwarz, \emph{Superstring Theory},
\emph{Phys.\ Rept.\ }{\bf 89}(1982), 233.

\bibitem{Green1}
M.B. Green, J.H. Schwarz, and E.~Witten, \emph{Superstring theory},
  vol.\ 1: \emph{Introduction}, Cambridge monographs on mathematical physics, Cambridge, 1987.

\bibitem{DeRoo1}
M.~De~Roo and M.G.C. Eenink, \emph{The effective action for the
4-point functions in abelian open superstring theory},
\emph{JHEP} {\bf 03}(08) 036 [{\tt hep-th/0307211}].

\bibitem{Chandia1}
O. Chandia and R. Medina, \emph{4-point effective actions in open
and closed superstring theory}, \emph{JHEP} {\bf 03}(11) 003
[{\tt hep-th/0310015}].

\bibitem{Brandt1}
F. Brandt, F. Machado and R. Medina, \emph{5-point effective
actions in open and closed superstring theory}, \emph{JHEP} {\bf
02}(07) 071 [{\tt hep-th/0208121}].

\bibitem{Kitazawa1}
Y.~Kitazawa, \emph{Effective lagrangian for the open superstring
from a 5-point function}, \emph{Nucl. Phys.} {\bf B 289}(1987), 599.

\bibitem{Refolli1}
A.~Refolli, A.~Santambrogio, N.~Terzi and D.~Zanon, \emph{$F^5$
  contributions to the non-abelian Born-Infeld action from a
  supersymmetric Yang-Mills five-point function}, \emph{Nucl. Phys.} {\bf B 613}(2001), 64
  [{\tt hep-th/0105277}]; Erratum-ibid: \emph{Nucl. Phys.} {\bf B 648}(2003), 453.

\bibitem{Koerber1}
P.~Koerber and A.~Sevrin, \emph{The non-abelian Born-Infeld action
  through order ${\alpha'}^3$}, \emph{JHEP} {\bf 01}(10) 003
  [{\tt hep-th/0108169 }]; \emph{Testing the ${\alpha'}^3$ term in the
  non-abelian open superstring effective action}, \emph{JHEP} {\bf 01}(09) 009
  [{\tt hep-th/0109030}].

\bibitem{Machado1}
F. Machado and R. Medina, \emph{The open superstring and the
non-abelian Born Infeld theory}, \emph{Nucl. Phys.}  {\bf B} \
(Proc. Suppl.) {\bf 127}  (2004), 166.

\bibitem{Barreiro1}
L. A. Barreiro and R. Medina, \emph{5-field terms in the open superstring
effective action}, \emph{JHEP} {\bf 05}(03) 055
[{\tt hep-th/0503182}].

\bibitem{Cederwall1}
M. Cederwall, B.E.W. Nilsson and D. Tsimpis,
\emph{D=10 super Yang-Mills at ${\alpha'}^2$},
\emph{JHEP} {\bf 01}(07) 042 [{\tt hep-th/0104236}].

\bibitem{Barreiro2}
L. A. Barreiro and R. Medina, work in progress.

\bibitem{Barreiro3}
L. A. Barreiro and R. Medina, work in progress.

\bibitem{Stieberger1}
D. Oprisa and S. Stieberger, \emph{Six gluon open superstring disk amplitude, multiple hypergeometric series and Euler-Zagier sums}, [{\tt hep-th/0509042}].

\bibitem{Koerber2}
P.~Koerber and A.~Sevrin, \emph{The non-abelian D-brane effective
action through order ${\alpha'}^4$}, \emph{JHEP}{\bf 02}(10) 042
[{\tt hep-th/0208044}].

\bibitem{Kawai1}
H. Kawai, D. C. Lewellen and S. -H. Tye, \emph{A relation between tree amplitudes of closed and open strings}, \emph{Nucl. Phys.} {\bf B 269}(1986), 1.

\end{thebibliography}
\end{document}